\documentclass{article}
\usepackage[letterpaper, total={6in, 8in}]{geometry}
\usepackage[utf8]{inputenc}
\usepackage{natbib}
\usepackage{graphicx}
\usepackage{xr}
\usepackage{authblk}
\usepackage[table,xcdraw]{xcolor}

\title{Accelerated Design of Block Copolymers: An Unbiased Exploration Strategy via Fusion of Molecular Dynamics Simulations and Machine Learning}

\author[1]{Jan Michael Y. Carrillo}
\author[2]{Vijith P}
\author[3]{Tarak K. Patra}
\author[4]{Zhan Chen}
\author[4,5]{Thomas P. Russell}
\author[6,7]{Subramanian KRS Sankaranarayanan}
\author[1,*]{Bobby G. Sumpter}
\author[2,6,*]{Rohit Batra}

\affil[1]{Center for Nanophase Materials Sciences, Oak Ridge National Laboratory, Oak Ridge,
Tennessee 37831, United States}
\affil[2]{Department of Metallurgical and Materials Engineering, Indian Institute of Technology Madras, Chennai 600036, India}
\affil[3]{Department of Chemical Engineering, Indian Institute of Technology Madras, Chennai 600036, India}
\affil[4]{Polymer Science and Engineering Department, Conte Center for Polymer Research, University of Massachusetts, Amherst, MA 01003, United States}
\affil[5]{Materials Sciences Division, Lawrence Berkeley National Laboratory, Berkeley, CA 94720, United States}
\affil[6]{Center for Nanoscale Materials, Argonne National Laboratory, Lemont, Illinois 60439, USA}
\affil[7]{Department of Mechanical and Industrial Engineering, University of Illinois, Chicago, Illinois 60607, USA}

\affil[*]{Author to whom correspondence should be addressed.}

\makeatletter
\newcommand*{\addFileDependency}[1]{
  \typeout{(#1)}
  \@addtofilelist{#1}
  \IfFileExists{#1}{}{\typeout{No file #1.}}
}
\makeatother

\newcommand*{\myexternaldocument}[1]{%
    \externaldocument{#1}%
    \addFileDependency{#1.tex}%
    \addFileDependency{#1.aux}%
}

\myexternaldocument{si}

\begin{document}
\maketitle

\begin{abstract}
    Star block copolymers (s-BCPs) have potential applications as novel surfactants or amphiphiles for emulsification, compatbilization, chemical transformations and separations. s-BCPs are star-shaped macromolecules comprised of linear chains of different chemical blocks (e.g., solvophilic and solvophobic blocks) that are covalently joined at one junction point. Various parameters of these macromolecules can be tuned to obtain desired surface properties, including the number of arms, composition of the arms, and the degree-of-polymerization of the blocks (or the length of the arm). This makes identification of the optimal s-BCP design highly non-trivial as the total number of plausible s-BCPs architectures is experimentally or computationally intractable. In this work, we use molecular dynamics (MD) simulations coupled with reinforcement learning based Monte Carlo tree search (MCTS) to identify s-BCPs designs that minimize the interfacial tension between polar and non-polar solvents. We first validate the MCTS approach for design of small- and medium-sized s-BCPs, and then use it to efficiently identify sequences of copolymer blocks for large-sized s-BCPs. The structural origins of interfacial tension in these systems are also identified using the configurations obtained from MD simulations. Chemical insights on the arrangement of copolymer blocks that promote lower interfacial tension were mined using machine learning (ML) techniques. Overall, this work provides an efficient approach to solve design problems via fusion of simulations and ML and provide important groundwork for future experimental investigation of s-BCPs sequences for various applications.
\end{abstract}

\hfill \break

\newpage

\section{Introduction}

Star block copolymers (s-BCPs) are an upcoming class of materials that can act as novel surfactants or amphiphiles for emulsification, compatbilization, chemical transformations and separations \cite{ren2016star}. As the name suggests, s-BCPs have an interesting star-shaped molecular architecture consisting of several polymer chains that are covalently linked at a single junction point. Depending on the chemical identity of the constituting polymer chains, the s-BCPs could either be regular or homo-arm type with several chemically identical chains of block copolymers, or hetero-arm or Mikto-arm type containing a variety of species with different chemical compositions, molecular weight or functionality \cite{liu2022miktoarm}. The Mikto-arm type star polymers are particularly interesting as their properties can be tuned by varying the design parameters. Furthermore, the s-BCPs have a limited tendency to aggregate and form micelles as they display high critical micelle concentration values, especially in comparison to their linear counterparts \cite{riess2003micellization,chou2008atypical,voulgaris1998polystyrene,mountrichas2005micelles}. Thus, they remain unimolecularly dispersed in a medium that is miscible with the corona block \cite{sheng2006morphologies,he2007synthesis,strandman2007self,zhang2018star}. Attempts have been made to exploit this property of s-BCPs to reduce interfacial tension ($\gamma_p$) between two immiscible liquids under dilute polymer/surfactant concentration limits
\cite{chen2013emerging,jerome1979star,xie2017heterografted,li2012peo}.



\begin{figure}[ht!]
\centering
\includegraphics[width=1\textwidth]{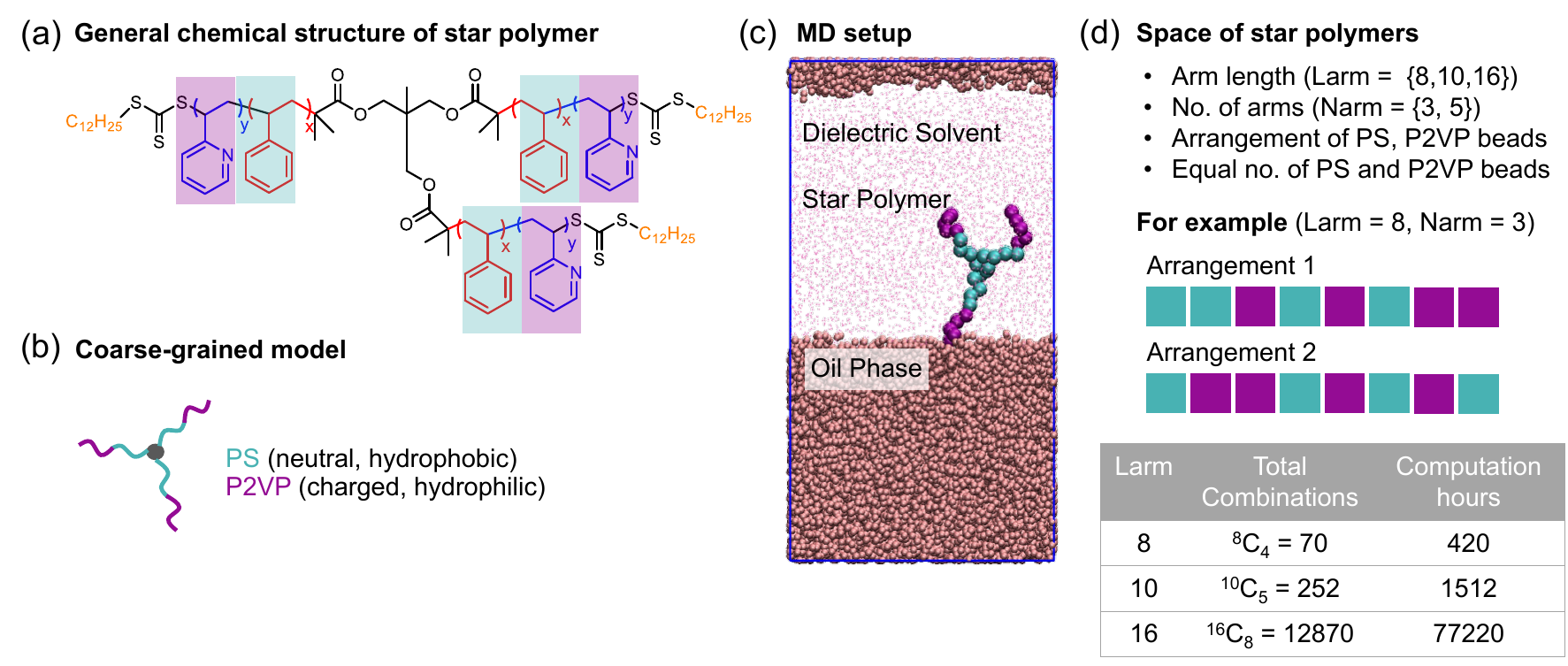}
\label{fig:Figure1}
\caption{\textbf{Chemical structure of s-BCPs and their potential design space.} (a) General chemical structure of a star polymer consisting of sequence of PS and P2VP. (b) A coarse-grained representation of a star polymer with three arms. (c) Schematic of the simulation cell consisting of a dielectric solvent, an oil phase and one star polymer. MD simulations were used to see the effect of s-BCP design on the interfacial tension ($\gamma_p$) between dielectric solvent and the oil phase. (d) The search space of the star polymer (and the associated computation time) considered in this work to identify sequence with low $\gamma_p$ values.}
\end{figure}

Figure \ref{fig:Figure1}(a) shows the general chemical structure of a s-BCP consisting of three copolymer chains with each containing a non-polar hydrophobic polystyrene (PS) block and a polar hydrophilic poly(2-vinylpyridine) (P2VP) block. We denote this s-BCP as (PS-$b$-P2VP)$_3$. At a low pH of $\sim$1.86 pKa, P2VP is mostly ionized with a positive charge and thus is hydrophilic \cite{roach2016counterion}. PS, on the other hand, is hydrophobic. Hence for a system containing oil/water interface, the (PS-$b$-P2VP)$_3$ can show interesting behavior with PS preferring the oil while the P2VP preferring the water phase.

Previous experimental and computational investigations have shown that linear block copolymers and s-BCPs can lower $\gamma_p$ of the oil/water interface \cite{fernandez2019interactions, carrillo2022assembly}. In our previous work \cite{carrillo2022assembly}, we used coarse-grained molecular dynamics (MD) simulations to study the assembly of star polymers at the oil/water interface (see Figures \ref{fig:Figure1}(b) and (c)),
wherein we considered a particular s-BCPs design that had the hydrophobic PS block concentrated at the junction point while the P2VP block forming the corona. On equilibration, (PS-$b$-P2VP)$_3$ adopted a specific conformation where the hydrophobic PS core resided in the oil phase with the P2VP corona extending into the aqueous phase. This was found to lower $\gamma_p$ between polar and non-polar solvents computationally, and was experimentally validated using tensiometry and vibrational sum frequency generation spectroscopy. It was also found the conformation adopted by the s-BCPs at the interface was highly dependent on the ionization degree of P2VP, the arm numbers and the molecular weight.


In this work we go beyond the previous studies in understanding the equilibrium conformation of (PS-$b$-P2VP)$_3$ star polymer (with a PS core and a P2VP corona) at the oil/water interface, and tackle the problem of star polymer design, as an s-BCP architecture with different arm sequence could potentially be a better surfactant. However, s-BCPs pose an intractable design challenge because of the large number of architectural possibilities. 
To realize a star polymer with desirable properties various parameters of these macromolecules need to be tuned, including the number of arms ($n_{\textrm{arm}}$), length of each arm ($l_{\textrm{arm}}$) (or degree-of-polymerization), and the relative composition and arrangement of the blocks within an arm. Figure \ref{fig:Figure1}(d) captures the plausible design space of star polymers assuming they consist of only two monomer types, a non-polar hydrophobic PS block and a quaternized (positively charged) polar hydrophilic P2VP block. We additionally constrain the number of arms, $n_{\textrm{arm}}$, to 2, 3 or 5; the length of each arm, $l_{\textrm{arm}}$, to 8, 10 or 16; and hold the ratio of PS to P2VP blocks to 1. Furthermore, we consider only the case of regular s-BCPs with concentration below the critical micelle concentration. Even with these constraints, there is a large number of design possibilities. For instance, for the case of $n_{\textrm{arm}} = 3$ and $l_{\textrm{arm}} = 8$, the relative arrangement of PS and P2VP beads can result in a total of $^{8}C_{4} = 70$ possible sequences---see Figure \ref{fig:Figure1}(d) for a few example sequence arrangements. To experimentally find the optimal star polymer design from such a large design space is infeasible. Computations could be used to explore this space in a brute force manner---an estimated total of $\sim$420 CPU hours ($\sim$6 hours for each MD simulation) would be needed to model all possible arrangements \cite{carrillo2022assembly}. However, as we approach more practical systems with larger values of $l_{\textrm{arm}}$, the brute force computational search becomes impractical as it requires 1512 and 77220 CPU hours for $l_{\textrm{arm}} = 10$ and $16$ systems, respectively.

To overcome these limitations of the brute search approach for star polymer design, we use reinforcement learning based Monte Carlo tree search (MCTS) coupled with coarse grained MD simulations. For the target property, we focus on identifying star polymers that can lower the $\gamma_p$ between polar and non-polar solvents, as has been considered in the past work \cite{carrillo2022assembly}. The MCTS algorithm balances the exploitation-vs-exploration trade-off to suggest promising star polymer designs that are expected to lower $\gamma_p$, while the MD simulations are used to evaluate the MCTS recommended designs and provide an accurate feedback to enhance the search quality. Further improvements are incorporated in the MCTS algorithm, such as the addition of design uniqueness criteria and an on-the-fly random forest surrogate model, to boost its performance. We first demonstrate the s-BCP design acceleration provided by our ML approach for a small ($l_{\textrm{arm}} = 8$) and a medium ($l_{\textrm{arm}} = 10$) sized space against the brute force and random search. Then we use our ML approach for a much larger ($l_{\textrm{arm}} = 16$) design space, which cannot be explored using MD simulations alone, to find new star polymers that can further lower $\gamma_p$. Structural analysis of various MD trajectories was performed to find correlations between arrangement of star polymer beads and $\gamma_p$. An additional random forest (RF) surrogate model was used to directly predict $\gamma_p$ of s-BCPs, and its features were analyzed to find important copolymer sequence arrangements that can result in lower $\gamma_p$. Overall, insights of this work could guide future experiments on s-BCP design for various applications.

\section{Results and Discussion}

\subsection{Small- and medium-sized star polymer space}
\begin{figure}[h!]
\centering
\includegraphics[width=1\textwidth]{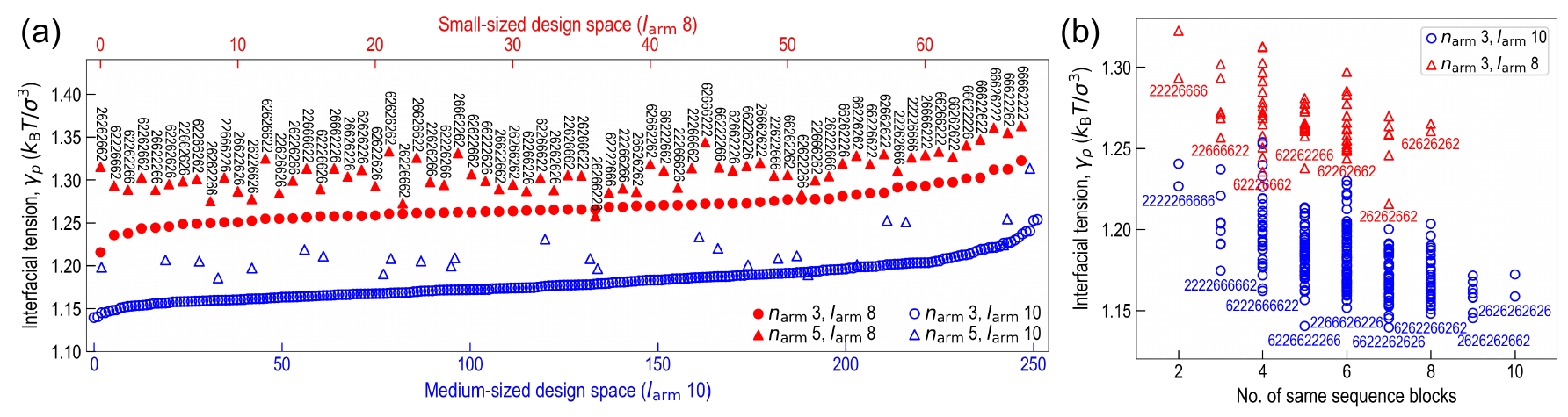}
\caption{\textbf{Interfacial energies of s-BCPs as computed using MD simulations.} (a) MD computed interfacial tension ($\gamma_p$) for the small ($l_{\textrm{arm}} = 8$) and medium ($l_{\textrm{arm}} = 10$) sized search space. The s-BCP sequences for $l_{\textrm{arm}} = 8$ are also included. While $\gamma_p$ increased with the increase in the number of arms from 3 to 5, it was not affected by the PS-P2VP sequence in a clear pattern. The order of s-BCP sequences on the x-axis is obtained based on increasing $\gamma_p$ values when $n_{\textrm{arm}} = 3$. (b) Trend in the $\gamma_p$ with increase in the number of same sequence blocks. Numbers 2 and 6 represent PS and P2VP beads, respectively.}
\label{fig:Figure2}
\end{figure}

MD simulations were performed to study the effect of $n_{\textrm{arm}}$ for small ($l_{\textrm{arm}} = 8$) and medium ($l_{\textrm{arm}} = 10$) sized star polymer design space. The details of the MD simulations and the simulated oil, water and star polymer system are provided in the Methods section. The difference between the normal and tangential pressures in the simulated system were used to compute the $\gamma_p$ between the oil/water interface as given by the equation \cite{tolman1948consideration, kirkwood1949statistical, ismail2006capillary}:

\begin{equation}
	\gamma_p = \frac{L_z}{2} \left<P_{zz}-\frac{P_{xx}+P_{yy}}{2}\right>
\label{eq:surface_tension}
\end{equation}

\noindent where $L_z$ is the simulation box dimension in the $z$ direction normal to the interface, $\left<...\right >$ refers to the ensemble-time average and the outer factor $\frac {1}{2}$ accounts for the two interfaces.

For $l_{\textrm{arm}} = 8$, three values of $n_{\textrm{arm}} = \{2,3,5\}$ were considered resulting in a total of 210 simulations. The obtained $\gamma_p$ values for $n_{\textrm{arm}} = \{3,5\}$ along with the associated PS-$b$-P2VP arrangement pattern are shown in Figure \ref{fig:Figure2}(a), while that for $n_{\textrm{arm}} = 2$ is included in the Supporting Information (Figure S3). It is important to note that previous experimental studies have shown that linear block copolymers (or when $n_{\textrm{arm}} = 2$) do not assemble at the interface but instead phase separate. For brevity, PS and P2VP beads are represented by numbers ``2" and ``6", respectively. Further, the first number in the s-BCP sequence represents the node center. Cases with a larger number of arms, in general, were found to have a higher $\gamma_p$. More importantly, candidates that contain longer and separate blocks of PS and P2VP beads (e.g., 66662222, 22226666 and 66622262) 
showed consistently high values of $\gamma_p$. However, no such trend was evident for cases with lower interfacial energies. Some example of sequences with the low $\gamma_p$ for the small-sized s-BCP design space with $n_{\textrm{arm}} = 3$ are 26262662, 62262266 and 26622266. Some of these designs have the PS bead at the junction point, while some have the P2VP bead. Some have alternating PS and P2VP beads, some have alternating blocks of multiple PS and P2VP beads. A general pattern in sequences with low $\gamma_p$ is absent.

Similar to the small-sized design space, MD simulations were performed for medium-sized space with $l_{\textrm{arm}} = 10$. Since the results from small-sized design space suggested lower $\gamma_p$ values for cases with a smaller number of arms, we exhaustively modeled all possible 252 sequences with only $n_{\textrm{arm}} = 3$, and about 30 sequences for validation purposes with $n_{\textrm{arm}} = 5$. As seen in Figure \ref{fig:Figure2}(a) similar trends were observed even for the medium-sized design space---cases with smaller number of arms resulted in lower interfacial energies, and a clear pattern among PS-P2VP sequences with lower $\gamma_p$ values was lacking. We note that a comparison between $\gamma_p$ values of medium and small-sized design spaces should not be made as each had a different number of total polymer beads, causing a change in chemical potential of the star polymer. However, such a comparison could be made across different values of $n_{\textrm{arm}}$ as care was taken to keep the number of beads constant.

Figure \ref{fig:Figure2}(b) shows the correlation between the computed $\gamma_p$ and the number of \textit{same} sequence blocks present in the star polymer. The results are shown for two cases with $n_{\textrm{arm}} = 3$, and $l_{\textrm{arm}} = 8$ or $10$. Here the number of same sequence blocks means that moving across the sequence how many times does a flip from 2 to 6 or from 6 to 2 occur. For example, for sequence 22226666 and 66662222 the flip occurs only once and the number of sequence blocks is 2 in each case. Similarly for sequence 26262626 and 22662266 the number of sequence blocks is 8 and 4, respectively. Figure \ref{fig:Figure2}(b) suggests that as the number of sequence blocks increases, $\gamma_p$ decreases, but the trend is not very strong. Towards the extreme right of the plot, where there is a large number of sequence blocks, the $\gamma_p$ values are moderately low, but not the lowest. Furthermore, there is huge variation in $\gamma_p$ values for the same number of sequence blocks. It is not very clear what causes the sequences 26262662 and 6622262626 respectively for $l_{\textrm{arm}} = 8$ and $10$, to exhibit low $\gamma_p$ values. This makes the design of star polymer non-trivial and highlights the need for an intelligent s-BCP design approach, especially for the large-sized design space which cannot be studied exhaustively.

\subsection{Structural origins of low energy sequences}
\begin{figure}[h!]
\centering
\includegraphics[width=1\textwidth]{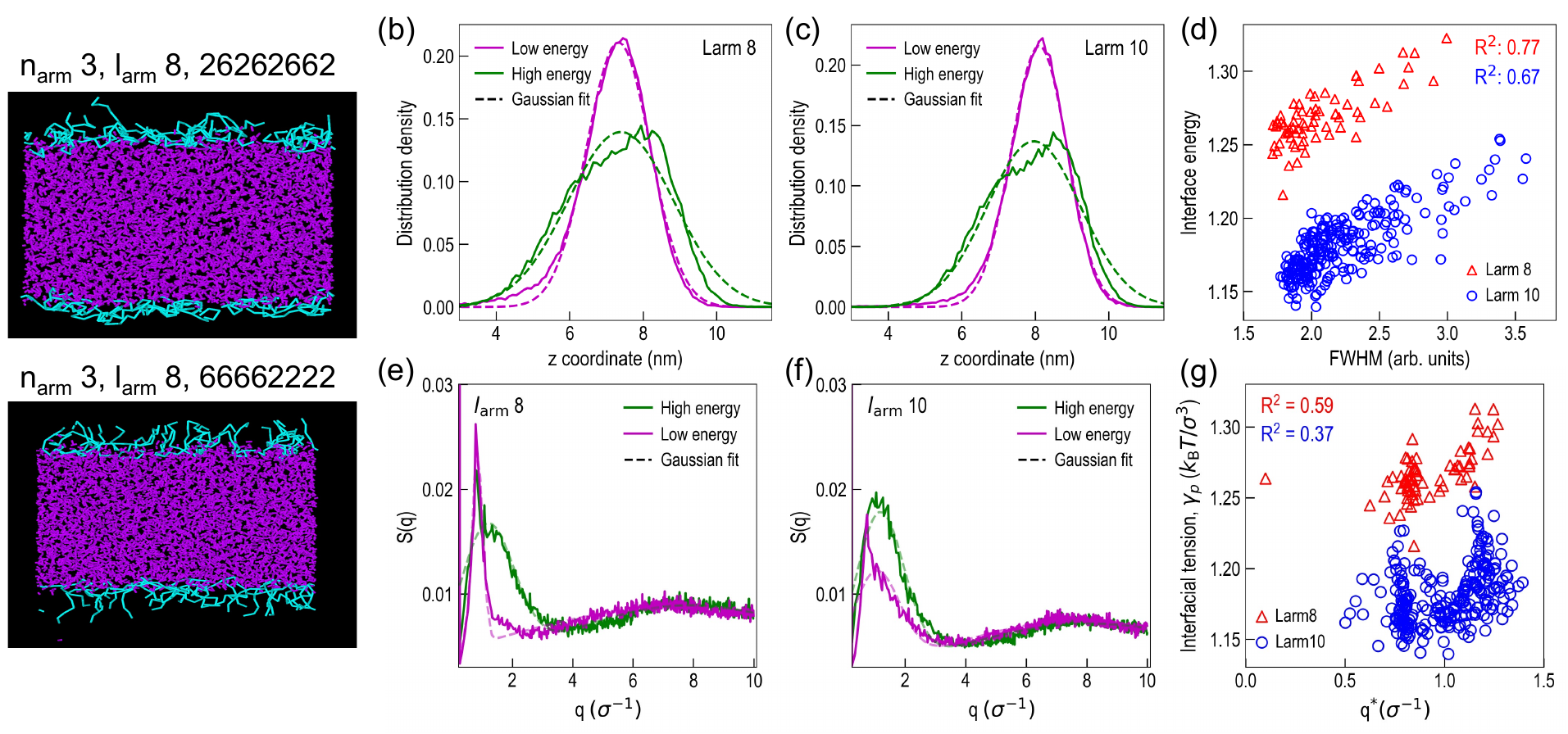}
\caption{\textbf{Structure-property correlations in s-BCPs.} (a) Output structures of the MD simulations for the lowest (top) and highest (bottom) interfacial tension ($\gamma_p$) values for design space with $l_{\textrm{arm}} = 8$ and $n_{\textrm{arm}} = 3$. Cyan and magenta colors represent the star polymer and oil beads, respectively; for clarity water beads are omitted. Systems with lower $\gamma_p$ formed sharper interface with the star polymer concentrated mostly at the interface, as shown in (b) and (c). (c) Strong correlation between the distribution of the star polymer in the direction normal to the interface and the computed $\gamma_p$ value for small- and medium-sized design space. Panels (e) and (f) show the distribution of PS beads in the lateral direction at the interface for the sequences with lowest and highest $\gamma_p$. (g) Weak correlations between $\gamma_p$ and the inter-molecular distance between PS beads in the lateral direction.}
\label{fig:Figure3}
\end{figure}

An assessment of the structural origins of the $\gamma_p$ values for different s-BCP designs was conducted. The structures corresponding to the lowest and the highest $\gamma_p$ values for the case with $l_{\textrm{arm}} = 8$ and $n_{\textrm{arm}} = 3$ are included in Figure \ref{fig:Figure3}(a). The density of the star polymer beads along the z-axis for these two cases is shown in Figure \ref{fig:Figure3}(b), along with the corresponding Gaussian fit line. From both these figures it is evident that the star polymer is more concentrated at the oil/water interface for the low energy case as compared to that in the high energy case; this is visually notable in Figure \ref{fig:Figure3}(a). The narrower and taller peak for the low energy case in Figure \ref{fig:Figure3}(b) further corroborates this. A similar observation of a narrower and taller star density peak for the case with $l_{\textrm{arm}} = 10$ and $n_{\textrm{arm}} = 3$ can also be made in Figure \ref{fig:Figure3}(c). This maybe due to availability of larger number of anchoring points at interface for the s-BCP designs with more alternating sequences.
To further quantify these observations, in Figure \ref{fig:Figure3}(d) we plot the computed $\gamma_p$ values against the full width at half maximum (FWHM) of the Gaussian fit of the star polymer density. A good correlation (0.67 for $l_{\textrm{arm}} = 8$ and 0.77 for $l_{\textrm{arm}} = 10$) between $\gamma_p$ and the FWHM values clearly suggests that the more the star polymers are concentrated at the interface, the lower is $\gamma_p$. Thus, the star polymer design should be such that a balance is maintained in its solubility in both the polar or non-polar solvents. If solubility is higher for either, the concentration peak will be broader and the expected $\gamma_p$ value higher. This also partly explains the earlier observation that sequences of alternating PS and P2VP blocks resulted in lower $\gamma_p$ values.   

We also analyzed the effect of the distribution of PS beads at the interface, i.e., in the lateral ($xy$) plane, on $\gamma_p$. For this, the 2D scattering function, $S(\vec{q})$ was computed for the PS beads as discussed in our previous work \cite{carrillo2022assembly}. In brief, we take the Fourier transform of the surface concentration of PS
beads in the $xy$ plane, whose magnitude at a $\vec{q}$ is obtained by taking product with its complex conjugate. Then, $S(\vec{q})$ is reduced to $S(q)$ using $q=\sqrt{q^{2}_x + q^{2}_y}$. Figure \ref{fig:Figure3}(e) and (f) shows the $S(q)$ values for the s-BCP sequences with lowest and highest $\gamma_p$ in the small- and medium-sized spaces, and $n_{\textrm{arm}} = 3$. Different peaks in reciprocal or $q$ space, corresponds to PS-PS bond lengths in real space with peaks at large $q$ values corresponding to shorter bonds. Generally, three distinct peaks were observed corresponding to PS-PS bonds that are inter-molecular, intra-molecular but in different arms, and intra-molecular within the same arm, as shown in Figures \ref{fig:Figure3}(e) and (f). The inter-molecular PS-PS bond peak is expected to occur at lowest $q$ value, labeled as $q^*$, and was analyzed further to explain the observed $\gamma_p$ behavior. A cumulative sum of three Gaussians was used to fit $S(q)$ data and the mean of the Gaussian around the lowest $q$ value was used as the $q^*$ value. Figure \ref{fig:Figure3}(g) captures the weak correlation between the computed $q^*$ and the $\gamma_p$ values, suggesting that the lateral arrangement of PS beads across different star polymer molecules does not significantly affect the interfacial tension at the oil/water interface. Overall, the above structural analysis suggests that arrangement of the polymers in the direction perpendicular to the interface is more important than that within the plane. This result could be useful in guiding experimental conformational studies of these systems.

\subsection{Efficient star polymer search using Monte Carlo tree search}

The small- and medium-sized s-BCP design space are amenable to brute force computational exploration, but not the large-sized design space. Thus, in this work we developed an ML approach for s-BCP design, which consists of MCTS algorithm and an MD simulator. Below we provide a brief description of our approach (see Methods section for details on the MCTS algorithm and MD simulator). While the MCTS performs the search for a promising (low $\gamma_p$ value) star polymer sequence in a tree-structured fashion, the MD simulator provides a feedback about the s-BCP design to the MCTS to further refine its search. Each leaf or node of the MCTS tree contains essential parameters describing a unique s-BCP design (such as  $l_{\textrm{arm}}$ and $n_{\textrm{arm}}$) and its associated score. The key idea behind MCTS is to efficiently sample the overall s-BCP design space by growing those branches of the tree that have either high scores (exploitation) or contain s-BCP designs that are very diverse (exploration) from the existing nodes in the tree. To provide a meaningful structure to the MCTS tree, each child node contains an s-BCP design which is only slightly different from that of the parent node. This results in high scoring child nodes generally belonging to a tree branch that contains other high scoring parent nodes with similar s-BCP designs.

\begin{figure}[h!]
\centering
\includegraphics[width=0.9\textwidth]{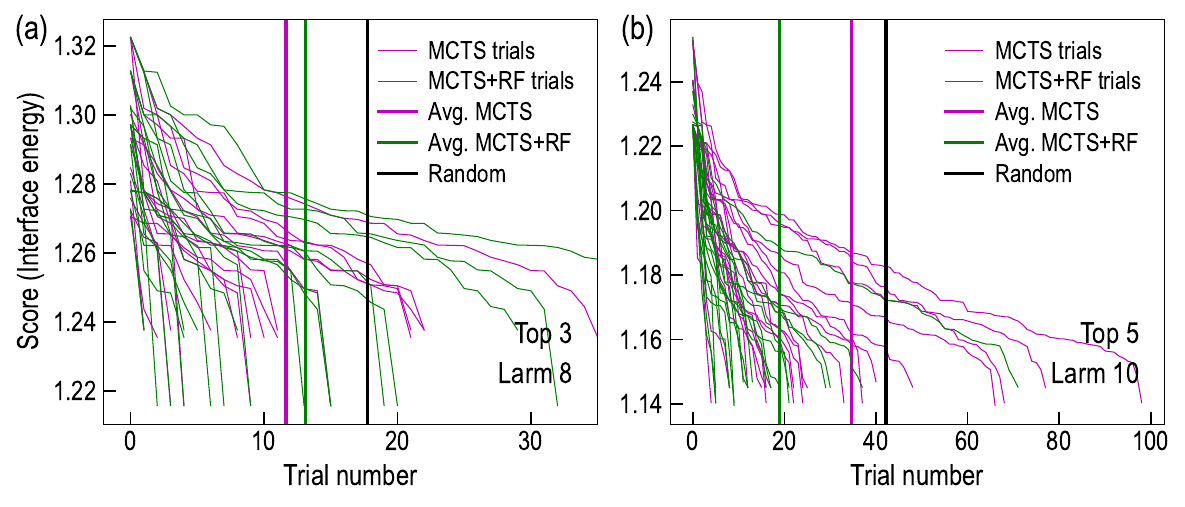}
\caption{\textbf{Comparative study of different ML models.} Comparison of the different methods to efficiently find the low interface energy star polymer sequences in the (a) small and (b) medium search space. MCTS and MCTS+RF schemes find one of the three (or five) lowest interface energy sequences in the small (or medium) size design space, on average, quicker than a random search. Results are shown for 20 random trails in each case.}
\label{fig:Figure4}
\end{figure}

MCTS uses two policies, namely a tree policy and a rollout policy, to advance the search. The upper confidence bound (UCB) for parameters is a popular choice of tree policy which have been used in the past \cite{liu2016modification}. Here, we use an improved version of the UCB policy where the exploration term is modified to promote uniqueness in the sampled candidates. Incorporation of the uniqueness criteria has been shown to improve the performance of the MCTS search for various materials discovery problems, such as peptide design \cite{batra2022machine}, fitting classical potentials \cite{manna2022learning}, and structure search \cite{banik2021learning}. For the rollout policy, two variations were adopted: one with just random rollouts, and the other one with half random rollouts and half based on a surrogate random forest (RF) model. The main idea behind the use of the RF model is to reduce the computation time spent on MD simulations, especially for the cases which are expected to have high $\gamma_p$. This could, however, introduce a bias in the search as the RF model itself is not accurate. Thus, to avoid the inherent bias of the approximate RF model only half of the rollouts were selected this way, while the remaining half were selected randomly. The RF model was trained in an online fashion, meaning that the RF model is regularly updated as more training data from the MD simulations become available during the MCTS run. It should be noted that once the s-BCP design is selected for a rollout, whether using random or RF model based rollout, its actual $\gamma_p$ value is computed using the MD simulator.

We first consider the small- and medium-sized design spaces to validate our ML approach for s-BCPs design. Both these spaces can be studied exhaustively using computations, and hence, these could be used to accurately evaluate the performance of our ML approach. Furthermore, we restrict our analysis to the candidates with $n_{\textrm{arm}} = 3$ since those resulted in low $\gamma_p$ values. Figure \ref{fig:Figure4} compares the number of trials needed for the various approaches to suggest any one of the top 3 and 5 star polymers designs for the small- and the medium-sized design spaces, respectively. The average number of trials needed for either of the MCTS or the RF model boosted MCTS approach (labelled, MCTS+RF) is smaller in comparison to the random search, highlighting the search acceleration provided by our ML approach. For instance, for the case with $l_{\textrm{arm}} = 10$, MCTS and MCTS+RF approaches respectively took on an average $\sim$34 and $\sim$19 trials as compared to $\sim$42 trials for the random search. Surprisingly, not much difference was found between the MCTS and MCTS+RF model for the small-sized design space. This is in contrast to our previous work where we clearly saw the performance boost provided by the RF model \cite{batra2022machine}. We hypothesize that this could be due to the small size of this design space. It should be noted that the performance gain of our ML approach is higher for the medium-sized design space than the small-sized one. This is expected since both the MCTS and the RF model need a few training examples to learn about the design space and propose high quality candidates. Thus, as the s-BCPs design space becomes larger the performance gain of our ML approach is expected to be higher. Another aspect to be noted in Figure \ref{fig:Figure4} is that MCTS approach has an inherent uncertainty associated with its performance; some trials were finished early, while some trials took many searches. Nevertheless, from these results we can conclude that our ML approach could be used to efficiently search for desirable s-BCP designs. 

\subsection{ML identified star polymers from large search space}

\begin{figure}[h!]
\centering
\includegraphics[width=0.9\textwidth]{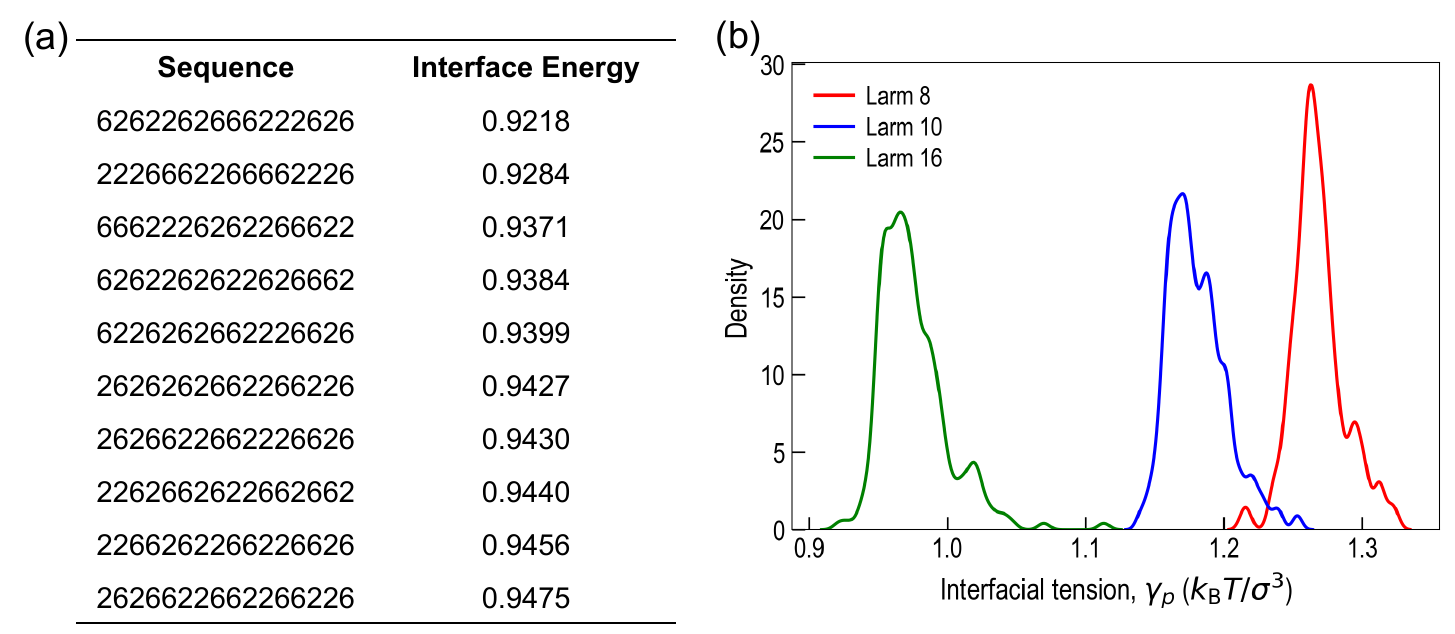}
\caption{\textbf{ML identified top s-BCP sequences.} (a) Top 10 sequences found in the large search space (Larm 16) using MCTS+RF search. (b) Similarity in the distribution of the interface energy found in the complete small (Larm 8) and medium (Larm 10), and, the partially explored, large (Larm 16) search space.}
\label{fig:Figure5}
\end{figure}

With the performance of the ML approach validated for the small- and medium-sized design spaces, we next use it to efficiently identify s-BCPs designs with low $\gamma_p$ values in the large-sized ($l_{\textrm{arm}} = 16$) design space. Figure \ref{fig:Figure5}(a) lists the top candidates and their computed $\gamma_p$ values identified by the ML approach from a total of 230 candidate evaluations. Again, no clear pattern of PS-P2VP arrangements could be observed from this list, highlighting the difficulty of this task. It should be noted that many candidates were discovered with lower $\gamma_p$ values than that of the candidate with alternating PS-P2VP beads, which was at fifteenth position in the list. This finding is similar to that observed in Figure 2 where candidates with the highest number of sequence blocks (or effectively alternating PS-P2VP beads) had relatively lower $\gamma_p$, but not the lowest. This list could be used to guide future s-BCP designs that could lower $\gamma_p$ between polar and non-polar solvents.

The large-sized design space cannot be explored exhaustively, as it has a total of 12870 possible candidates. Thus, to make an assessment of the quality of the top candidates identified by the ML approach, we plot the distribution of $\gamma_p$ values obtained from all the three design spaces in Figure \ref{fig:Figure5}(b). The similarity in the shape of the three design spaces provides some confidence that the ML approach has effectively explored the large-sized design space. We also include the structural analysis of the evaluated 230 candidates in the Supporting Information (Figure S2) which was found to be similar to that of Figure \ref{fig:Figure3}(c).

\subsection{Chemical insights from the ML models}
\begin{figure}[h!]
\centering
\includegraphics[width=1\textwidth]{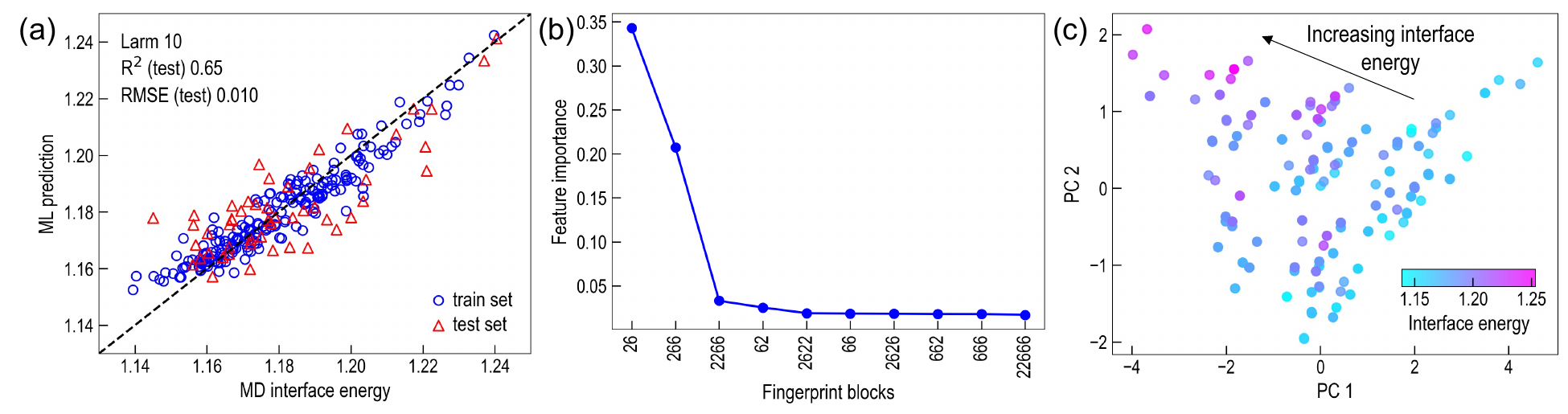}
\caption{\textbf{Chemical insights from ML models.} (a) Accuracy of the RF machine learning model trained to predict the interface energy associated with the star polymer using the medium-sized design space. (b) Top 10 most important features found from the RF model. (c) Projection of the medium space dataset on the first two principal components using only the top 10 features identified by the RF model.}
\label{fig:Figure6}
\end{figure}

We further wanted to understand which type of sequence blocks are responsible for lowering $\gamma_p$ and if any chemical insights could be mined from the generated MD data. For this, we begin by training a separate RF model that directly predicts the $\gamma_p$ of a candidate given its arrangement of PS and P2VP beads. This model was restricted to the dataset with $l_{\textrm{arm}} = 10$ and $n_{\textrm{arm}} = 3$. The input fingerprint for this model was the count of different types of possible single (2), double (4), triple (8), quadruple (16) and pentuple (32) block sequences where the number in the bracket indicate the feature count of a particular category. A few example features from the different categories are ``6", ``22", ``662", ``2266" and ``22226". The overall fingerprint dimension was 62. The good performance accuracy (coefficient of determination, R$^2 = 0.65$) achieved by the RF model on the test set in Figure \ref{fig:Figure6}(a) is indicative of the success of this fingerprinting scheme, and thus this model was next analyzed to gain reliable chemical insights.

Figure \ref{fig:Figure6}(b) plots the feature importance of the top 10 sequence blocks according to the RF model. Two features, namely the count of ``26" and ``266" blocks, have significantly higher importance when compared to the rest. To further understand this, we studied the correlation plots between the computed $\gamma_p$ values and the suggested top 5 RF model features (see Supporting Information, Figure S4). A strong negative correlation between the count of ``26" blocks and $\gamma_p$ values was observed. This can be understood from the previous observation that the candidates with alternating PS-P2VP blocks, which also means higher count of ``26" block, result in lower $\gamma_p$ values. A weak negative correlation with feature ``266" was also found suggesting that simply alternating between PS-P2VP beads is not sufficient but a complex sequence arrangement is needed to reach the lowest $\gamma_p$ s-BCP design. Figure \ref{fig:Figure6}(c) presents the medium-sized design space data along the first two principal components obtained using only the aforementioned top 10 features. The colors indicate the $\gamma_p$ values. A clear trend in regards to the spatial arrangement of the points with similar $\gamma_p$ values in same places suggest that these top 10 features are indeed able to capture the complex interdependence of these features to affect the $\gamma_p$ values.

\section{Conclusions}
In this work we studied various star copolymer configurations of PS and P2VP that could minimize the interfacial tension between polar and non-polar solvents. We studied the effect of number of star copolymer arms, the length of each arm and the sequence arrangement of PS and P2VP blocks using coarse-grained molecular dynamics simulations. While an increase in number of star copolymer arms were found to increase the interfacial tension, an inverse trend was found with the arm length. More interestingly, no clear trend of interfacial tension with the sequence of PS and P2VP in the star copolymer was observed, thus making the problem of identifying the lowest interfacial tension copolymer sequences non-trivial, especially when the length of copolymer arm is large and an exhaustive sequence search is not possible.

To efficiently search for low interfacial tension sequences, we employed a reinforcement learning based Monte Carlo tree search, which was further accelerated using a uniqueness criteria in the scoring function and an on-the-fly random forest surrogate model. To validate the improvement gain of our machine learning approach, we first employed it on a smaller and medium sized copolymer search space, wherein the ground truth can be established using an exhaustive search, and compared its performance against the random search. Next, we used our approach to find top scoring sequences in the large sized copolymer search space wherein the ground truth cannot be uncovered due to computational costs.

Structural origins of interfacial tension were also identified using the configurations obtained from the molecular dynamics simulations. As expected, star polymer sequences that have high concentration at the interface, demonstrate lower interfacial tension. 
We also extracted chemical insights from the random forest model to reveal the count of ``PS-P2VP" blocks in the star copolymer inversely co-related to the interfacial tension. Other important chemical block sequences that dominate this interfacial tension problem were also extracted. Overall, this work provides an efficient approach to solve design problems using machine learning and enables important groundwork for future experimental investigation of star copolymer sequences that could lower interfacial tension between polar and non-polar solvents.

\section{Methods}
\subsection{Molecular dynamics simulations}
Coarse-grained MD simulations were performed to probe the effect of star diblock copolymers on the interfacial tension of the dielectric solvent/oil interface. The simulated system consists of three main components: the star diblcok copolymers, the dielectric solvent phase, and the oil phase. The star block copolymers were represented as connected coarse-grained beads denoting neutral PS segments, positively charged P2VP segments and explicitly added negatively charged counterions. All short-range pair-wise interactions are described by the shifted truncated Lennerd-Jones (LJ) potential as discussed in more detail in the Supporting Information. The bond connectives were described by the finite extensible non-linear elastic (FENE) model with spring constant $k_{bond}$ (see Supporting Information).

The next component in the model is the dielectric solvent which is represented as charged dumbbells having 2 opposite charges with a magnitude of $q$ and separated by some distance. The pairwise interactions of dielectric solvent beads include both short-range LJ interactions and long-range Coulomb interaction. Finally, the oil phase is represented as LJ beads where the oil and solvent is incompatible, PS is miscible to oil and P2VP is slightly less miscible to oil beads relative to PS beads. The pair-wise non-bonded potential parameters are summarized in Table S1. The simulation consisted of 5000 solvent molecules (10000 solvent beads) and 5000 oil beads, while the number of PS/P2VP beads were 480, 600 and 960 for small-, medium- and large-sized s-BCP design spaces, respectively. Additional details on the different model parameters are provided in the Supporting Information.

The simulation consists of an initial equilibration isothermal-isobaric ensemble (NPT)  step followed by a canonical (NVT) production run.
The temperature was maintained by coupling the system to the Langevin thermostat \cite{schneider1978molecular}. A Berendsen barostat \cite{berendsen1984molecular} was used to control pressure in the equilibration stage. During the NPT simulation, the dimension of the simulation box is adjusted to achieve the target pressure via coupling the barostat to the $x$, $y$ directions. During the production NVT runs, the simulation box is fixed to the average dimensions determined in the NPT runs. The NPT equilibration run proceeded for up to  $1.5\times10^4\;\tau$ and the NVT production run proceeded for up to $1.0\times10^4\;\tau$. The velocity-Verlet algorithm with a time step of $\Delta t = 0.005$ $\tau$ was used for integrating the equations of motion in Eq. S5.  All simulations were performed using the LAMMPS molecular dynamics simulations software package \cite{plimpton1995fast,brown2011implementing}.

\subsection{Monte Carlo tree search}
Monte Carlo tree search (MCTS) is a powerful global optimization algorithm owing to its exploration-versus-exploitation trade-off and low computational demand \cite{coulom2006efficient, kocsis2006bandit, browne2012survey}. It has been particularly successful in solving problems involving large search spaces \cite{silver2016mastering, dieb2019monte, srinivasan2021artificial}. In this work, we use MCTS to suggest promising s-BCP designs with low $\gamma_p$ values. MCTS performs the search in a tree-structure manner wherein each leaf or node of a tree contains a unique s-BCP design. Moreover, leaf nodes contain connections in a special configuration such that a parent node is connected to several child nodes with slightly different s-BCP designs. Thus, each branch of the MCTS tree contains similar s-BCP designs.

The MCTS search iterates through following four stages: (1) \textit{selection}: select the leaf node that has the highest current score according to the tree policy; (2) \textit{expansion}: add a child node (with slightly different s-BCP design from the parent node) to the selected leaf node; (3) \textit{simulation}: perform Monte Carlo trials of possible actions on the newly generated child node using a rollout policy to estimate the expected reward; (4) \textit{back-propagation}: pass the rewards generated by the simulated trials to update the scores of all the parent leaves of the newly generated child node. An important distinction between score and reward should be made. The former is computed using the tree policy given by \cite{liu2016modification}:

\begin{equation}
\mathrm{UCB}(\theta_{j}) = - \mathrm{min}(r_{1}, r_{2},...,r_{n_{i}}) + c.f(\theta_{j}).\sqrt{\frac{ln N_{i}}{n_{i}}}
\label{eq:ucb}
\end{equation}

\noindent where $\theta_{j}$ represents the node $j$ in the MCTS tree, $r$ denotes the reward ($\gamma_p$ value) of a given rollout, $c (> 0)$ is the exploration constant, $n_i$ is the number of rollout samples taken by node $\theta_{j}$ and all of its child nodes, and $N_i$ is the same value as $n_{i}$ except for the parent node of $\theta_{j}$. On the other hand, reward represents the property that we are trying to maximize/minimize, i.e., $\gamma_p$ value in this work. The second term on the right in Eq. \ref{eq:ucb} represents the exploration part and promotes the search in those regions of the design space that have not been investigated yet.


In this work, the MCTS was initiated by setting a random s-BCP design to the root node at depth 0. The number of Monte Carlo trials were set to 2 for small and medium-sized, and 8 for large-sized design spaces. In the scenario when no RF model was used, all of the Monte Carlo trials were performed on randomly generated sequences. In contrast, within the MCTS+RF scheme half of the trials were performed on randomly generated sequences, while the remaining half were screened from a pool of 10 random designs based on RF model predicted reward. The exploration constant, $c$, was set to 10 for the large-sized design space, while it was optimized for the small- and medium-sized design space using grid search (see Supporting Information, Figure S5). In order to train the RF model and to compute the uniqueness function, a fingerprinting function was devised to numerically represent a s-BCP design, as described in section 2.5. 
The uniqueness function $f(\theta_j)$ was computed as the mean of the cosine distances of s-BCP design fingerprints to that of all other designs in the tree.

\subsection{Random forest model and principal component analysis}
The random forest (RF) regression algorithm, as implemented in the scikit-learn \cite{scikitlearn}, was used to learn the $\gamma_p$ values of the s-BCP designs. The RF is based on the concept of ensemble learning, wherein predictions from several ``weak" models are averaged to overall result in a better prediction accuracy.  Thus, RF is expected to perform better than other ML methods based on just a single model construction, such as Gaussian process regression. Besides accuracy, the training time of RF model is also not prohibitively long, especially with large training datasets. This makes RF algorithm suitable for on-the-fly training during MCTS runs. The RF hyperparameters, i.e., the number of weak estimators and the maximum depth of the tree, were determined using cross-validation. It should be noted that two types of RF models were trained: one, for the MCTS runs, and the other to mine new chemical insights. 
For both the models, the count of different PS-P2VP sequence blocks was used as the model input, as discussed in section 2.5. The RF model was trained to minimize the mean squared error. Principal component analysis included in Figure \ref{fig:Figure6}(c) was performed using the PCA python library, as implemented in the scikit-learn \cite{scikitlearn}.

\section{Data Availability} 
The MD computed $\gamma_p$ data for all s-BCP designs is available as Supporting Information.

\section{Code Availability} 
The machine learning code used in this work is available as Supporting Information.

\section{Competing Interests}
The authors declare no competing financial or non-financial interests.

\section{Author Contributions}
J.M.Y.C, S.K.R.S.S, B.G.S. and R.B. conceived and designed the project. J.M.Y.C set up the MD simulations. R.B. designed and executed the ML algorithm for star polymer design. V.P. analyzed the MD simulation results and evaluated the performance gain of the ML algorithm. All the authors contributed to the data analysis and to the
preparation of the manuscript. J.M.Y.C, S.K.R.S.S, B.G.S. and R.B. wrote the manuscript with input from
all the coauthors. B.G.S. and R.B. supervised and directed the overall project.

\bibliographystyle{plain}

\providecommand{\latin}[1]{#1}
\makeatletter
\providecommand{\doi}
{\begingroup\let\do\@makeother\dospecials
	\catcode`\{=1 \catcode`\}=2 \doi@aux}
\providecommand{\doi@aux}[1]{\endgroup\texttt{#1}}
\makeatother
\providecommand*\mcitethebibliography{\thebibliography}
\csname @ifundefined\endcsname{endmcitethebibliography}
{\let\endmcitethebibliography\endthebibliography}{}

\section{Acknowledgements}
A part of this research was performed at the Center for Nanophase Materials Sciences (CNMS) at Oak Ridge National Laboratory and the Center for Nanoscale Materials (CNM) at Argonne National Laboratory, which are US Department of Energy (DOE) Office of Science User Facilities. MD simulations used resources of the Oak Ridge Leadership Computing Facility, which is supported by DOE Office of Science under Contract DE-AC05-00OR22725. The overall concept of using ML to accelerate materials discovery and to enable digital twins for soft matter was supported by the U.S. Department of Energy, Office of Science, Office of Basic Energy Sciences Data, Artificial Intelligence and Machine Learning at DOE Scientific User Facilities Program under Award Number 34532. RB acknowledges support from Center for Atomistic Modelling and Materials Design under the IOE scheme, ICSR, IIT Madras for the initiation research grants, and from Robert Bosch Centre for Data Science and AI (RBCDSAI).
\end{document}